\documentclass[draf]{aa}
\usepackage{natbib}
\usepackage{color}
\usepackage{graphicx}
\usepackage[varg]{txfonts}
\usepackage{hyperref}
\hypersetup{citecolor=black, urlcolor=blue, linkcolor=black, colorlinks=true}
\newcommand{\fig}[1]{Fig.\,\ref{#1}}
\newcommand{\figs}[1]{Figs.\,\ref{#1}}

\definecolor{orange}{rgb}{1,0.4,0.}
\graphicspath{{./}{figs-ps/}{figs/}}
\begin{document}

%=============================================================================
% TITLE
%=============================================================================
%
\title{Bidirectional propagating brightenings in arch filament systems observed by Solar Orbiter/EUI} 
\titlerunning{bidirectional brightenings in AFS}

%\authorrunning{}

\author{Yajie Chen \inst{1}
          \and
          Sudip Mandal \inst{1}
          \and
          Hardi Peter \inst{1}
          \and
          Lakshmi Pradeep Chitta \inst{1}
          }

\institute{
             \inst{1} Max-Planck Institute for Solar System Research, 37077 G\"{o}ttingen, Germany
             \email{cheny@mps.mpg.de}
          }

\date{Version: \today}
%\date{Received xx Jan 2021 / Accepted xx yyy 2021}

  \abstract
  {
  Arch filament systems (AFSs) are chromospheric and coronal manifestations of emerging magnetic flux.
  Using high spatial resolution observations taken at a high cadence by the Extreme Ultraviolet Imager (EUI) on board Solar Orbiter, we identified small-scale elongated brightenings within the AFSs.
  These brightenings appear as bidirectional flows along the threads of AFSs.
  For our study, we investigated the coordinated observations of the AFSs acquired by the EUI instrument and the Atmospheric Imaging Assembly (AIA) on board the Solar Dynamics Observatory (SDO) on 2022 March 4 and 17.
  We analyzed 15 bidirectional propagating brightenings from EUI 174 {\AA} images.
  These brightenings reached propagating speeds of 100--150 km~s$^{-1}$.
  The event observed on March 17 exhibits blob-like structures, which may be signatures of plasmoids and due to magnetic reconnection.
  In this case, we also observed counterparts in the running difference slit-jaw images in the 1400 {\AA} passbands taken by the Interface Region Imaging Spectrograph (IRIS).
  Most events show co-temporal intensity variations in all AIA EUV passbands. Together, this implies that these brightenings in the AFSs are dominated by emission from cool plasma with temperatures well below 1 MK. 
  The Polarimetric and Helioseismic Imager (PHI) on board Solar Orbiter provides photospheric magnetograms at a similar spatial resolution as EUI and from the same viewing angle.
  The magnetograms taken by PHI show signatures of flux emergence beneath the brightenings.
  This suggests that the events in the AFSs are triggered by magnetic reconnection that may occur between the newly emerging magnetic flux and the preexisting magnetic field structures in the middle of the AFSs.
  This would also give a natural explanation for the bidirectional propagation of the brightenings near the apex of the AFSs.
  The interaction of the preexisting field and the emerging flux may be important for mass and energy transfer within the AFSs.
  }

%
%----------------------------------------------------------------------------

\maketitle

%==============================================================================
\section{Introduction\label{S:intro}}
%==============================================================================

Arch filament systems (AFSs) form over magnetic polarity inversion lines (PILs) in active regions and connect magnetic patches of opposite polarity. They are manifestations of emerging magnetic flux \citep[e.g.,][]{Zwaan1985}
and appear as parallel dark fibrils in the chromosphere \citep[e.g.,][]{Bruzek1967}.
Simulations of flux emergence can reproduce many properties of AFSs, such as their flow patterns and filamentary structures \citep[e.g.,][]{Shibata1989,Fan2001,Isobe2006}.
Magnetic reconnection between the newly emerging magnetic flux associated with the AFSs and the preexisting magnetic field structures may produce extreme ultraviolet (EUV) brightenings \citep[e.g.,][]{Tarr2014}, coronal loops \citep[e.g.,][]{Longcope2005}, and flares \citep[e.g.,][]{Zuccarello2008,Su2018}.

Because of the absorption of hydrogen and helium continua, AFSs can be recognized as dark structures in coronal images in EUV wavelengths \citep[e.g.,][]{Mein2001}, and transient loop-like brightenings occasionally occur within AFSs \citep[e.g.,][]{Deng2000}.
Numerical simulations of filaments suggest that turbulent heating at their footpoints can produce elongated threads at coronal temperatures within the filaments through evaporation \citep{Zhou2020}.
Based on High-Resolution
Coronal Imager \citep[Hi-C;][]{Hi-C2.1} observations at the 172 {\AA} passband with a spatial resolution of $\sim$500 km,
\citet{Tiwari2019} found that the loop-like brightenings within an AFS originate from the footpoints. These brightenings are related to flux cancellation in the photosphere, inferring that the events are triggered by magnetic reconnection near the footpoints of the AFS.
{In addition, \citet{2019ApJ...887L...8P} reported jet-like events also from Hi-C 2.1 observations and proposed that these events are triggered by flux cancellation and associated mini-filament eruption.}

In two-dimensional scenarios, magnetic reconnection can produce bidirectional outflows \citep{2014masu.book.....P}.
The bidirectional flows have been revealed in spectroscopic observations of flares \citep[e.g.,][]{Hara2011} and explosive events \citep[e.g.,][]{Innes1997}, often serving as evidence for magnetic reconnection.
In addition, plasmoid instability can segment current sheets with high Lundquist numbers and is important for fast energy release during magnetic reconnection  \citep[e.g.,][]{1963PhFl....6..459F, 2000mrp..book.....B}.
Blob-like structures in solar observations are often explained as an observable proxy of plasmoids produced during fast magnetic reconnection,
and they have been found in various types of events \citep[e.g.,][]{Takasao2012,2016NatPh..12..847L,Kumar2018,2023A&A...673A..11R,Cheng2023,Hou2024,Cheng2024},
but observations of such blob-like structures are still limited due to their small scales.

The High Resolution Imager (HRI) of the Extreme Ultraviolet Imager \citep[EUI;][]{EUI} on board Solar Orbiter \citep{SolO} at the 174 {\AA} passband (HRI$_{\mathrm{EUV}}$) takes coronal observations with unprecedentedly high spatial and temporal resolution.
{Recent observations taken by HRI$_{\mathrm{EUV}}$ have revealed ubiquitous campfires \citep{2021A&A...656L...4B,2021ApJ...921L..20P,2021A&A...656A..35Z} in the quiet Sun regions and pico-flare jets \citep{2023Sci...381..867C} in the coronal holes.}
Additionally, HRI$_{\mathrm{EUV}}$ has also made it possible to resolve substructures in various types of small-scale events and capture their temporal evolution.
For example, bidirectional moving structures in small-scale loops \citep[e.g.,][]{Chitta2021,Li2022} and plasmoids in jets \citep{Mandal2022,Cheng2023,Li2023} have been seen in the images taken by HRI$_{\mathrm{EUV}}$.

Emission patterns from the cores of active regions with plasma temperatures in excess of 3\,MK\ \citep[e.g.,][]{2011ApJ...740..111T} are thought to be the signature of nanoflares \citep[][]{2011ApJ...738...24V}, which are a direct manifestation of magnetic reconnection \citep[][]{2015RSPTA.37340256K}. Thus, active region cores and structures within, including AFSs, are good observational targets to investigate the properties of blobs and further understanding of the spatial and temporal scales of magnetic reconnection operating in the corona.
In this paper, we report on bidirectional propagating brightenings in the AFSs in the core of the active regions observed by HRI$_{\mathrm{EUV}}$.
The HRI$_{\mathrm{EUV}}$ observations, with a cadence as high as 3\,s and a {pixel size of images} of $\sim$135\,km taken during the perihelion of Solar Orbiter, make it possible to resolve the blob-like structures within the brightening and study their dynamics.
Furthermore, we examine magnetic field maps in the photosphere to investigate the generation mechanisms of the loop-like brightenings.

%==============================================================================
\section{Observations \label{S:obs}}
%==============================================================================

%------------------------------------------------------------------------------
\subsection{2022 March 4}
%------------------------------------------------------------------------------

The first dataset was taken on 2022 March 4.
The distance between Solar Orbiter and the Sun was $\sim$0.53 AU, and the angle between the Sun-Earth line and the Sun-Solar Orbiter line was $\sim$4.6$^{\circ}$.
The HRI$_{\mathrm{EUV}}$ performed observations of NOAA AR 12957 during 10:45--11:45 UT.
The cadence of the EUI 174 {\AA} images was 5 s, and the image scale $\sim$189\,km\,pixel$^{-1}$.
The EUI dataset was taken from the Level 2 data of EUI Data Release 6.0 \citep{euidatarelease6}.
The High Resolution Telescope (HRT) of the Polarimetric and Helioseismic Imager \citep[PHI;][]{PHI} on board Solar Orbiter performed coordinated observations during 10:45--11:39\,UT with a cadence of 3\,minutes.
PHI/HRT took observations of the full Stokes profiles of the Fe\,{\sc i}\,6173\,\AA\ line at five wavelength positions along with a nearby continuum point.
Reduction and calibration of SO/PHI data are detailed in \citealt{2022SPIE12180E..3FK}, \citealt{2022SPIE12189E..1JS}, and \citealt{2023A&A...675A..61K}.
We only used the derived line-of-sight magnetograms.
The image scale of the magnetograms taken by PHI/HRT was $\sim$194\,km\,pixel$^{-1}$.

We also analyzed the coordinated observations of the Atmospheric Imaging Assembly \citep[AIA;][]{AIA} and the Helioseismic and Magnetic Imager \citep[HMI;][]{HMI} on board the Solar Dynamics Observatory \citep[SDO;][]{SDO}.
The cadences of the AIA images were 12\,s in the 94, 131, 171, 193, 211, 335, and 304 {\AA} passbands and 24\,s in the 1600 {\AA} passband.
The pixel size of the AIA images was $\sim$435 km.
The cadence of the magnetograms taken by HMI was 45 s, and the pixel size was $\sim$365 km.
Coalignment among the AIA images in different passbands and the HMI magnetograms was achieved by using the IDL routine \textit{aia\_prep.pro} in SolarSoft.
We then coaligned the HRI$_{\mathrm{EUV}}$ images and AIA images by comparing the common bright patterns in the HRI$_{\mathrm{EUV}}$ and the AIA 171 {\AA} images.
Furthermore, we compared the magnetograms obtained by PHI/HRT and HMI to coalign the observations taken by PHI/HRT and other instruments.

%------------------------------------------------------------------------------
\subsection{2022 March 17}
%------------------------------------------------------------------------------

The second dataset was taken on 2022 March 17.
The distance between Solar Orbiter and the Sun was $\sim$0.38 AU, and the angle between the Sun-Solar Orbiter line and the Sun-Earth line was $\sim$26.4$^{\circ}$.
HRI$_{\mathrm{EUV}}$ took images of NOAA AR 12965 during 03:18--04:03 UT with a cadence of 3 s, and the pixel size of the images was $\sim$135 km.
The EUI dataset was also taken from the Level 2 data of EUI Data Release 6.0 \citep{euidatarelease6}.
PHI/HRT performed coordinated observations during 03:18--03:47 UT with a cadence of 1 minute,
and the pixel size of the magnetogram was $\sim$138 km.
We used coordinated AIA and HMI observations to assist with the coalignment between EUI and PHI observations.
We first compared the AIA 171 {\AA} images and magnetograms taken by HMI, and then we compared the HRI$_{\mathrm{EUV}}$ images and magnetograms taken by PHI/HRT to match the correspondence.

The Interface Region Imaging Spectrograph \citep[IRIS;][]{IRIS} performed very large sit-and-stare observations between 03:23 UT and 03:57 UT.
The slit-jaw images (SJIs) were taken in 1400 and 2796 {\AA} passbands with a cadence of $\sim$3.6 s, and the spatial pixel size was $\sim$239 km.
When we were examining the responses of the HRI$_{\mathrm{EUV}}$ emission patterns in the IRIS observations, we first re-projected the HRI$_{\mathrm{EUV}}$ images onto the field of view of IRIS based on the World Coordinate System (WCS) keywords and then compared the emission patterns in SJI 1400 {\AA} and HRI$_{\mathrm{EUV}}$ images.
The coalignment is illustrated in Fig. D.1 of \citet{Mandal2023}.
Unfortunately, the slit did not cross the brightening analyzed in this study, so we did not analyze the spectral observations of IRIS.

%==============================================================================
\section{Results \label{S:results}}
%==============================================================================

%>>>>>>>>>>>>>>>>>>>>>>>>>>>>>>>>>>>>>>>>>>>>>>>>>>>>>>>>>>>>>>>>>>>>>>>>>>>>>>
\begin{figure*}[ht]
\centering {\includegraphics[width=14cm]{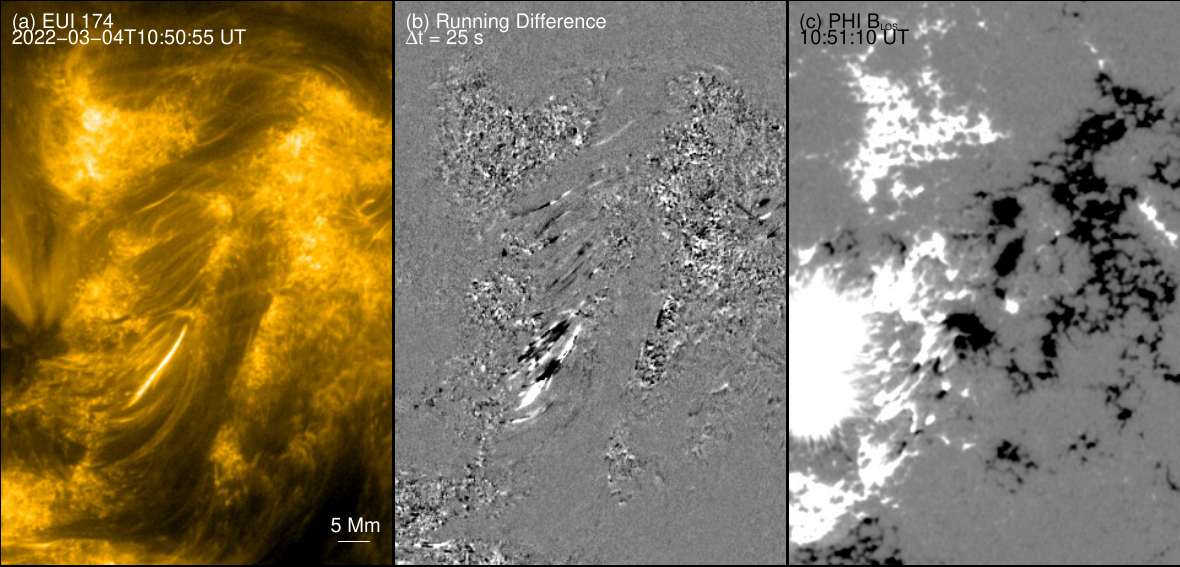}} 
\caption{
Observations of the arch filament system on 2022 March 4.
(a) HRI$_{\mathrm{EUV}}$ image taken at 10:50:55 UT. 
Only part of the field of view is shown.
(b) Intensity difference of the EUI 174 {\AA} images shown in (a) and the one taken at 10:50:30 UT. 
(c) Line-of-sight component of the photospheric magnetic field taken by PHI at 10:51:10 UT.
The HRI$_{\mathrm{EUV}}$ image is shown in a logarithm scale with arbitrary units, and the magnetogram is saturated at $\pm$350\,G.
{An animation is available online, and the events are highlighted by arrows.}
} 
\label{fig:overview1}
\end{figure*}
%<<<<<<<<<<<<<<<<<<<<<<<<<<<<<<<<<<<<<<<<<<<<<<<<<<<<<<<<<<<<<<<<<<<<<<<<<<<<<<

%>>>>>>>>>>>>>>>>>>>>>>>>>>>>>>>>>>>>>>>>>>>>>>>>>>>>>>>>>>>>>>>>>>>>>>>>>>>>>>
\begin{figure*}[ht]
\centering {\includegraphics[width=\textwidth]{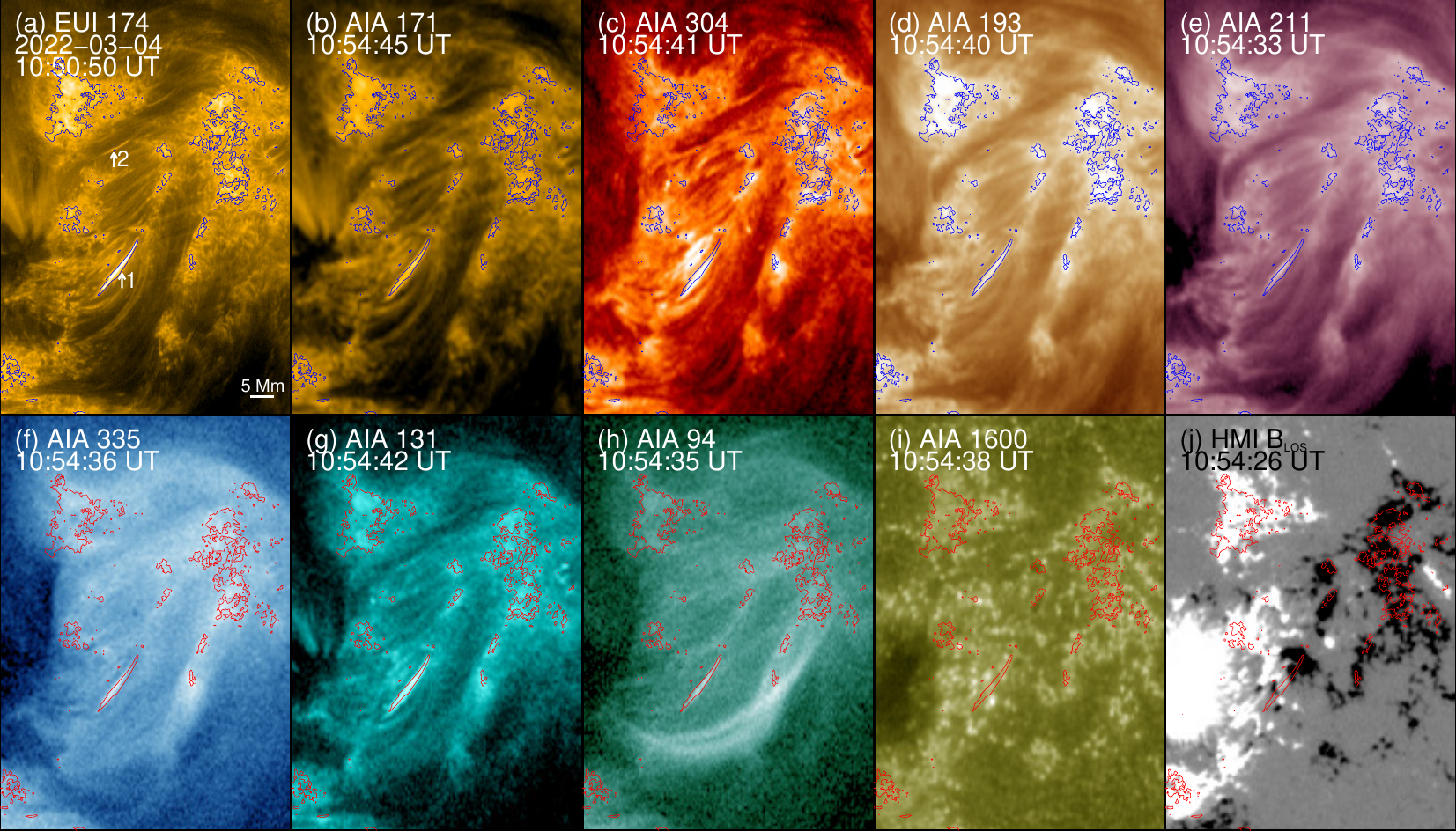}} 
\caption{
Coordinated observations of AIA and HMI taken on 2022 March 4.
(a) EUI 174 {\AA} image taken at 10:50:50 UT.
The two arrows indicate the two events shown in \figs{fig:case1} and \ref{fig:case2}.
(b--i) AIA 171, 304, 193, 211, 335, 131, 94, and 1600 {\AA} images taken around 10:54 UT, respectively.
(f) Magnetogram taken by HMI at 10:54:26 UT.
The EUI and AIA images are shown in logarithm scales with arbitrary units, and the magnetogram is saturated at $\pm$350 G.
The contours outline the regions with enhanced EUI 174 {\AA} emissions.
{An animation is available online, and the events are highlighted by arrows.}
} 
\label{fig:aia}
\end{figure*}
%<<<<<<<<<<<<<<<<<<<<<<<<<<<<<<<<<<<<<<<<<<<<<<<<<<<<<<<<<<<<<<<<<<<<<<<<<<<<<<

%>>>>>>>>>>>>>>>>>>>>>>>>>>>>>>>>>>>>>>>>>>>>>>>>>>>>>>>>>>>>>>>>>>>>>>>>>>>>>>
\begin{figure*}[ht]
\centering {\includegraphics[width=13cm]{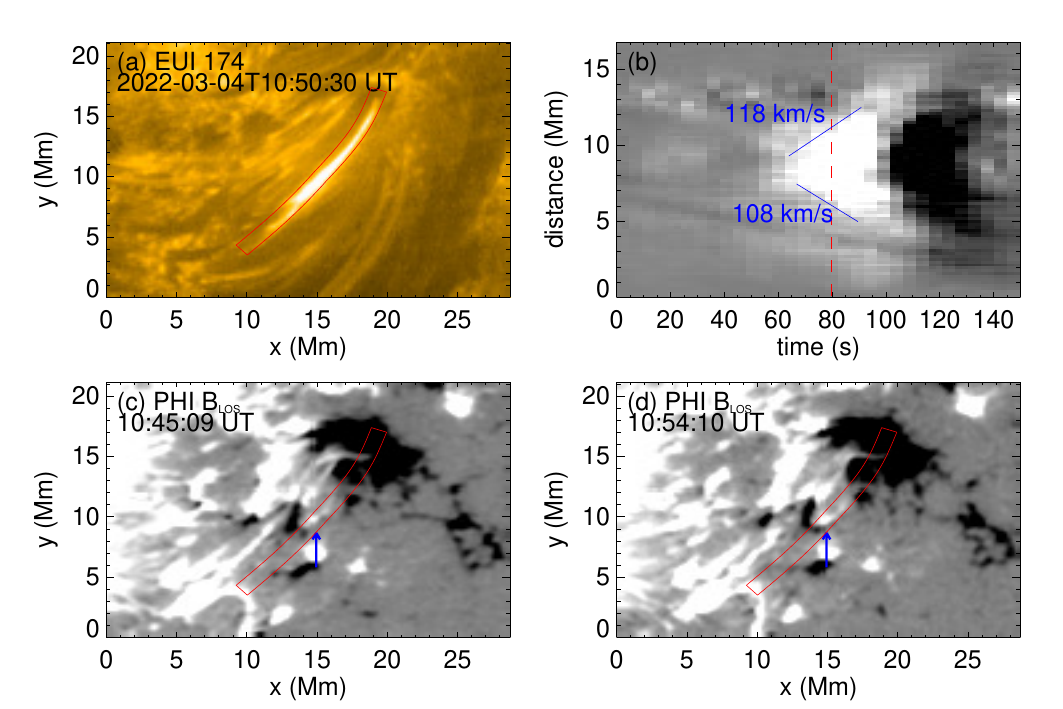}} 
\caption{
Coordinated observations of EUI and PHI of the first case observed on 2022 March 4.
(a) Zoomed-in EUI 174 {\AA} image taken at 10:50:30 on 2022 March 4.
(b) Space-time diagram of the running difference images in EUI 174 {\AA} passband along the slit shown in (a).
The speeds are estimated through the slope of the blue lines.
(c--d) Magnetograms saturated at $\pm$350 G taken by PHI at 10:48:10 and 11:03:10 UT, respectively.
{The blue arrows point out the location of flux emergence and cancellation.}
} 
\label{fig:case1}
\end{figure*}
%<<<<<<<<<<<<<<<<<<<<<<<<<<<<<<<<<<<<<<<<<<<<<<<<<<<<<<<<<<<<<<<<<<<<<<<<<<<<<<

%>>>>>>>>>>>>>>>>>>>>>>>>>>>>>>>>>>>>>>>>>>>>>>>>>>>>>>>>>>>>>>>>>>>>>>>>>>>>>>
\begin{figure*}[ht]
\centering {\includegraphics[width=12cm]{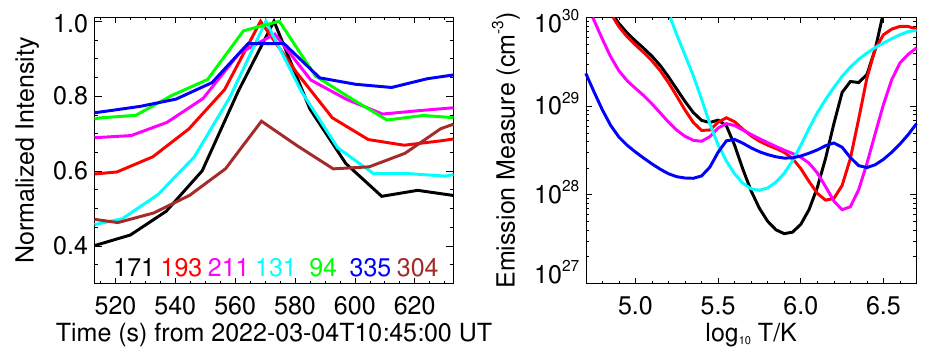}} 
\caption{
Light curves and EM-loci curves for the first case.
Left: Intensity variations within the region outlined by the red lines shown in \fig{fig:case1}(a).
The intensity for each passband is summed within the region outlined by the red curves in \fig{fig:case1} and normalized to the peak intensity of the passband. 
Right: EM-loci curves for the brightening shown in \fig{fig:case1}(a).
The different colored lines correspond to different AIA passbands.
} 
\label{fig:lc}
\end{figure*}
%<<<<<<<<<<<<<<<<<<<<<<<<<<<<<<<<<<<<<<<<<<<<<<<<<<<<<<<<<<<<<<<<<<<<<<<<<<<<<<

The AFS observed by HRI$_{\mathrm{EUV}}$ on 2022 March 4 is presented in \fig{fig:overview1}.
The magnetogram taken by PHI/HRT is also shown in \fig{fig:overview1}.
The AFS consists of many parallel dark fibrils connecting the large-scale positive and negative polarities in the active region.
Many bright loop-like brightenings appear between the dark fibrils.
Underneath the AFS, there are many small-scale magnetic patches with mixed polarities.
The bright lanes may be related to the evolution of small-scale magnetic field elements, which are discussed later.
From the running difference of the HRI$_{\mathrm{EUV}}$ images, as shown in \fig{fig:overview1}(b), we found elongated brightening features that appear as emission-enhanced regions and occur on either side of a relatively darker emission-depleted region.
This indicates that the brightening originates in the middle and propagates as a bidirectional flow.
Our aim is to study the properties of these bidirectionally propagating brightenings and their relationship to the magnetic activities in the photosphere.

To this end, we first examined the responses of the brightenings in the AIA observations as shown in \fig{fig:aia}.
AIA 171 {\AA} and HRI$_{\mathrm{EUV}}$ images are similar (see \fig{fig:aia}(a--b)), as their filter-response functions have peaks around similar temperatures ($\sim$10$^{5.9}$ K in AIA 171 {\AA} and $\sim$10$^{6.0}$ K in EUI 174 {\AA} passbands).
%
%This also indicates that the coalignment between EUI and AIA images are acceptable.
%
The spatial resolution of HRI$_{\mathrm{EUV}}$ is about two times higher than that of the AIA 171 {\AA} passband, and thus more details are resolved in the HRI$_{\mathrm{EUV}}$ images.
To investigate the dynamics of the brightening (arrow 1 in \fig{fig:aia}(a)), we applied an artificial slit as shown in \fig{fig:case1}(a). The width of the artificial slit is $\sim$9.5 Mm.
At each location of the brightening along the slit, we took the average intensity in each HRI$_{\mathrm{EUV}}$ image across the slit.
%
%Then we subtracted the intensity along the slit by that obtained from one frame earlier.
%
In this way we obtained the space-time diagram along the artificial slit.
Furthermore, we smoothed the space-time diagram over 3$\times$3 pixels and then subtracted it from the original space-time diagram in order to enhance the visibility of the moving feature.
The processed space-time diagram is shown in \fig{fig:case1}(b).
At this point, we could then determine the plane-of-sky component of the propagating speeds of the brightening edges as 108 and 118~km~s$^{-1}$, respectively.

Figure \ref{fig:aia}(c--h) shows that the elongated brightening in the HRI$_{\mathrm{EUV}}$ image also exhibits as a brightening in the AIA 304, 193, 211, 335, 131, and 94 {\AA} images.
We present the light curves of different AIA passbands in \fig{fig:lc} within the box outlined in \fig{fig:case1}(a).
The light curves for all EUV passbands of AIA peak at almost the same time, and the time lags among the different passbands are negligible. 
{Such thermal behavior is similar to the events studied in \citet{Winebarger2013} and \citet{Tiwari2019}.}
%
%and it implies that the plasma is most likely to be at temperatures less than a million degree Kelvin (i.e., not coronal), and the cool plasma dominates the intensity variations in different AIA passbands.
%
In such cases, it is challenging to determine the temperature distribution based on differential emission measure (DEM) analysis from AIA observations. This is because the AIA passbands poorly constrain the emission from cooler plasma at transition region temperatures \citep{DelZanna2011,Testa2012}. 
Therefore, we calculated the emission measure loci (EM-loci) curves at the center of the brightening in order to derive the possible temperature distribution \citep{DelZanna2002}.
The results are shown in \fig{fig:lc}, and the EM-loci curves of different AIA passbands cross at a narrow temperature range, $\sim$10$^{5.5}$ K.
It is worth mentioning that the temperature range of response functions of AIA passbands are wide, and such thermal analyses can have large uncertainties \citep[e.g.,][]{Tian2014}.

The line-of-sight components of the magnetic field maps in the photosphere taken by PHI/HRT before and after the event's appearance are shown in \fig{fig:case1}(c--d).
For this case, the surrounding magnetic patches in the HMI observations are also visible but with much less detail.
Although the cadence of PHI/HRT observations was four times poorer than HMI, PHI/HRT still captured the evolution of the surrounding magnetic field elements.
Below the center of the elongated brightening, the magnetic flux of each of the two small-scale opposite polarities becomes stronger, gets closer to the other, and then interacts with the other. %emerge and then interact with each other.
%
%Thus, it is possible that magnetic reconnection associated with flux emergence and cancellation in the photosphere triggered the event and generated bidirectional flows along the AFS.

%>>>>>>>>>>>>>>>>>>>>>>>>>>>>>>>>>>>>>>>>>>>>>>>>>>>>>>>>>>>>>>>>>>>>>>>>>>>>>>
\begin{figure*}[ht]
\centering {\includegraphics[width=13cm]{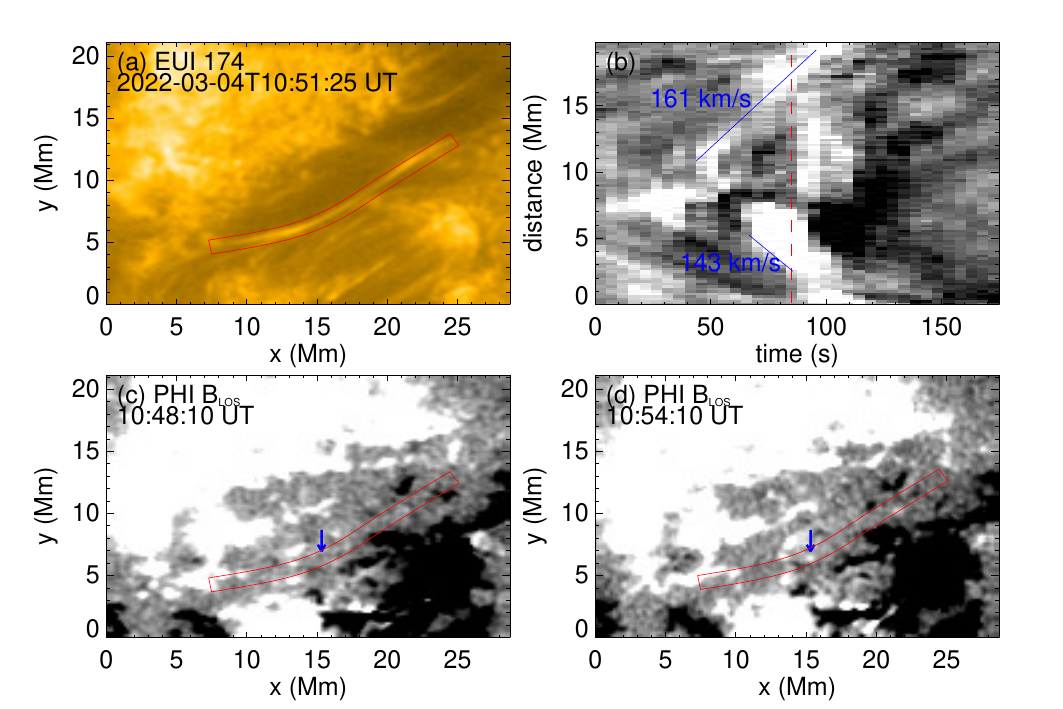}} 
\caption{
Similar to \fig{fig:case1} but for the second case observed on 2022 March 4.
Magnetograms are saturated at $\pm$40 G.
{The blue arrows in (c--d) point out the location of flux emergence.}
} 
\label{fig:case2}
\end{figure*}
%<<<<<<<<<<<<<<<<<<<<<<<<<<<<<<<<<<<<<<<<<<<<<<<<<<<<<<<<<<<<<<<<<<<<<<<<<<<<<<

Figure \ref{fig:case2} presents the second event identified from the first dataset (arrow 2 in \fig{fig:aia}).
We put an artificial slit with a width of $\sim$9.5\,Mm along the elongated brightening as shown in \fig{fig:case2}(a) and generated the space-time diagram as shown in \fig{fig:case2}(b) following the procedures mentioned above.
The estimated speeds of the brightenings are approximately $\sim$150~km~s$^{-1}$.
Moreover, we observed an emission enhancement in the AIA 304, 193, 211, and 131 {\AA} passbands at the same location.
There are no clear signatures in the AIA 335 and 94 {\AA} passbands, and this may be because there are some hot loops at $\sim10^{6.5}$ K above the event, and the emission from the loop makes the event indistinguishable in the 335 and 94 {\AA} images.
The magnetograms taken by PHI are also presented in \fig{fig:case2}(c--d). 
One magnetic patch pointed out by the arrow in \fig{fig:case2}(d) appears at $x\simeq15$ Mm below the brightening, which is a clear signature of flux emergence.
The bidirectional flows may be triggered by magnetic reconnection between the emerging flux and the overlying field lines of the AFS.
%
%It is worth mentioning that HMI does not have sufficient sensitivity and spatial resolution to resolve such newly emerging magnetic flux elements.
%
We further isolated another 12 events exhibiting bidirectional flows from the dataset taken on 2022 March 4,
and their speeds mostly reach 100--150~km~s$^{-1}$.
Additionally, all of them show signatures of flux emergence and/or cancellation in the magnetograms taken by PHI/HRT.

%>>>>>>>>>>>>>>>>>>>>>>>>>>>>>>>>>>>>>>>>>>>>>>>>>>>>>>>>>>>>>>>>>>>>>>>>>>>>>>
\begin{figure*}[ht]
\centering {\includegraphics[width=13cm]{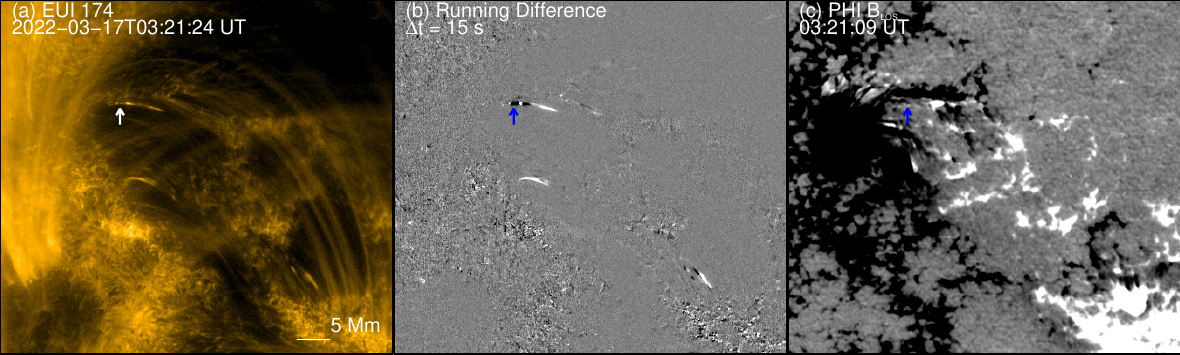}} 
\caption{
Similar to \fig{fig:overview1} but for the arch filament system taken on 2022 March 17.
The magnetogram is saturated at $\pm$150 G.
The arrows indicate the event shown in \fig{fig:case3}.
The time difference of the EUI 174 {\AA} images to obtain (b) is 15 s.
An animation is available online.
} 
\label{fig:overview2}
\end{figure*}
%<<<<<<<<<<<<<<<<<<<<<<<<<<<<<<<<<<<<<<<<<<<<<<<<<<<<<<<<<<<<<<<<<<<<<<<<<<<<<<

%>>>>>>>>>>>>>>>>>>>>>>>>>>>>>>>>>>>>>>>>>>>>>>>>>>>>>>>>>>>>>>>>>>>>>>>>>>>>>>
\begin{figure*}[ht]
\centering {\includegraphics[width=14cm]{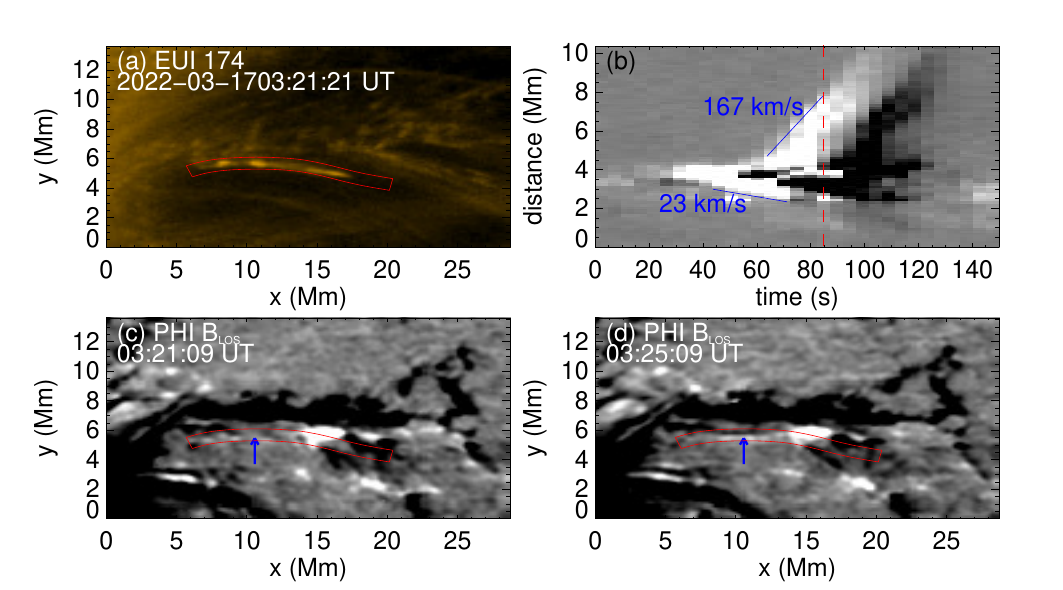}} 
\caption{
Similar to \fig{fig:case2} but for the third case observed on 2022 March 17.
{The blue arrows in (c--d) point out the location of flux emergence.}
} 
\label{fig:case3}
\end{figure*}
%<<<<<<<<<<<<<<<<<<<<<<<<<<<<<<<<<<<<<<<<<<<<<<<<<<<<<<<<<<<<<<<<<<<<<<<<<<<<<<

%>>>>>>>>>>>>>>>>>>>>>>>>>>>>>>>>>>>>>>>>>>>>>>>>>>>>>>>>>>>>>>>>>>>>>>>>>>>>>>
\begin{figure*}[ht]
\centering {\includegraphics[width=13cm]{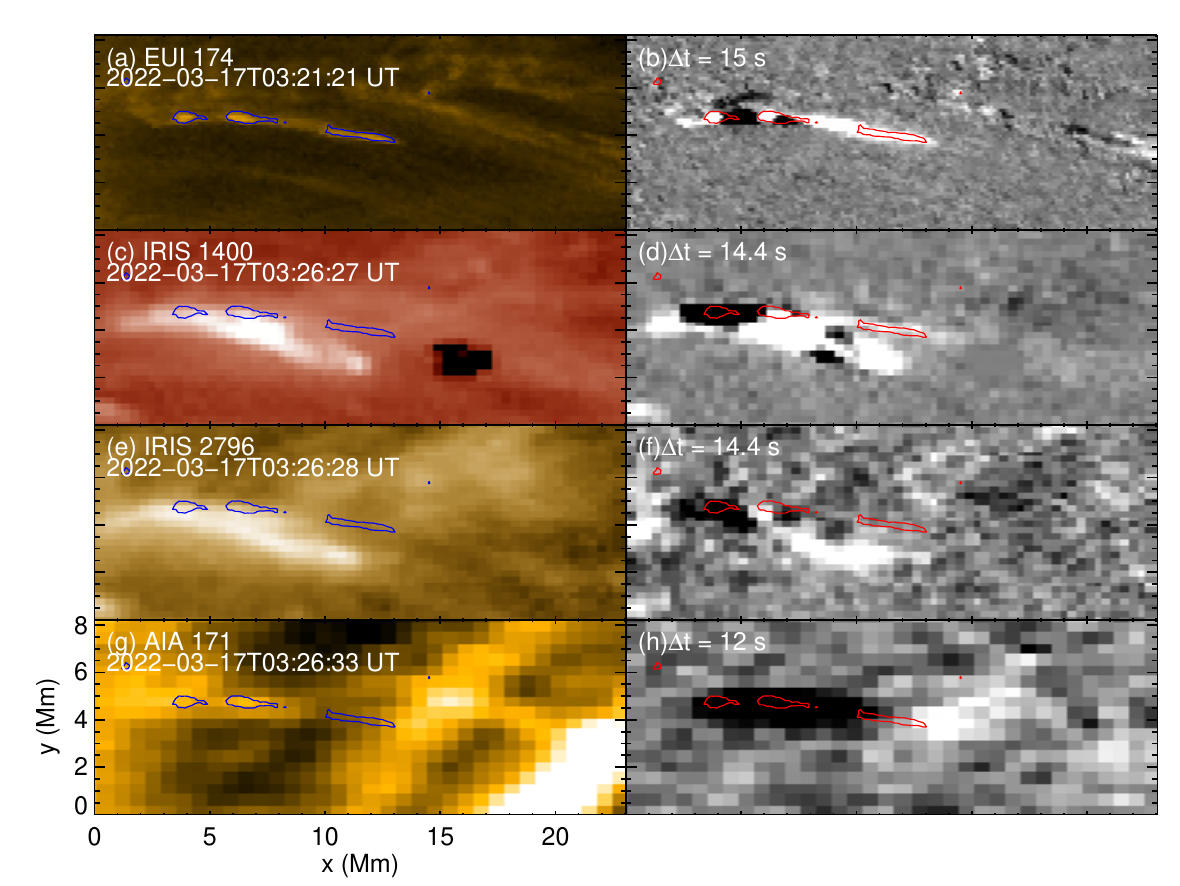}} 
\caption{
Coordinated IRIS observations of the third event taken on 2022 March 17.
(a) EUI 174 {\AA} image taken at 03:21:18 UT.
(b) Intensity difference between the images taken at 03:21:21 and 03:21:06 UT.
(c--d) Similar to (a--b) but for the IRIS 1400 {\AA} passband.
(e--f) Similar to (a--b) but for the IRIS 2796 {\AA} passband.
(g--h) Similar to (a--b) but for the AIA 171 {\AA} passband.
Contours outline the regions with enhanced EUI 174 {\AA} emission.
} 
\label{fig:eui_iris_aia}
\end{figure*}
%<<<<<<<<<<<<<<<<<<<<<<<<<<<<<<<<<<<<<<<<<<<<<<<<<<<<<<<<<<<<<<<<<<<<<<<<<<<<<<

The third example is taken from the observations obtained on 2022 March 17.
The dataset from March 17 does not have as many bright lanes as the one taken on March 4.
%
%We noticed that the small-scale magnetic field patches beneath the AFS are not as prevelant as the previous dataset, and the lower levels of the magnetic field activities may be associated with less brightenings in the AFS.
%
Part of the AFS in HRI$_{\mathrm{EUV}}$ is shown in \fig{fig:overview2}, and the magnetogram taken by PHI/HRT is also presented.
We focused on one example at the northernmost region of the AFS, and the zoomed-in field of view of the EUI 174 {\AA} image taken at 03:21:21 is shown in \fig{fig:case3}(a).
The brightening consists of many bright blobs, and such substructures are resolved likely because the spatial resolution of the EUI instrument increases as the Solar Orbiter spacecraft proceeds toward its perihelion.
We derived the space-time diagram along the artificial slit shown in \fig{fig:case3} (a), and the result is exhibited in \fig{fig:case3}(b).
The brightening reveals bidirectional flows along the slit.
The speed of the blobs at the west side is $\sim$ 167~km~s~$^{-1}$, which is consistent with the speeds of the cases in observations of March 4.
The speed of the blobs at the east side is $\sim$23~km~s~$^{-1}$.
{The event originates near the footpoint of the AFS, which could restrict plasma propagation toward the footpoint due to density stratification. As a result, the plasma flow in the two directions is highly asymmetric.}
{Besides, the blobs likely propagate mainly along the line-of-sight direction near the footpoint, and the derived plane-of-sky component of the velocity is low due to the projection effect.}
In order to understand the generation mechanism of the bidirectional propagating blobs, we examined the evolution of magnetograms below the event, as shown in \fig{fig:case3}(c--d).
{As pointed out with blue arrows}, magnetograms reveal signatures of flux emergence in the photosphere within the red box, and the newly emerging positive and negative polarities likely undergo shearing motions.
{This event is similar to the jets in the tiny loops in AFS observed by Hi-C 2.1 \citep[e.g.,][]{Tiwari2019,2019ApJ...887L...8P}, although the jets in their studies are mainly unidirectional flows. \citet{Tiwari2019} also found bidirectional flows in the AFS, but those were mostly small surges.}

Imaging observations of IRIS also captured the third example, so we examined the response of the bidirectional propagating blobs in the SJI 1400 and 2796 {\AA} images as shown in \fig{fig:eui_iris_aia}(c--f).
The bright lane in the EUI 174 {\AA} passband shows clear offsets compared to the bright loops in both the SJI 1400 and 2796 {\AA} images.
By further comparing the bright features in the AIA 171 {\AA} and SJI 1400 and 2796 {\AA} images, we found that they also show clear offset.
Thus, the offsets between the brightenings in coronal and transition region images are not because of the coalignment between EUI and IRIS based on the WCS keywords.
\citet{Peter2022} reported that the loops in IRIS and Hi-C observations exhibit offsets, and the loops in the transition region and corona are adjacent but not identical.
This might also be the case in our observations.
Although not clear in the original images, the running difference of the SJI 1400 {\AA} images presents clear enhancement at the locations of the bright blobs in the EUI 174 {\AA} images, as shown in \fig{fig:eui_iris_aia}(f).
There is no obvious response in the SJI 2796 {\AA} images, suggesting the blobs are heated above chromospheric temperatures.
Since nearby coronal loops overlap with the AFS in AIA observations and it is challenging to isolate the emission from the blobs and loops above, we did not perform further analyses of the AIA observations.

%==============================================================================
\section{Discussions\label{S:diss}}
%==============================================================================

In this study, we isolated several elongated brightenings propagating bidirectionally along the threads within AFSs from EUI 174 {\AA} images.
Most events observed on 2022 March 4 show identical morphology in all AIA EUV channels, and the intensity variations in these passbands match well temporally.
Thus, these brightenings may be dominated by emission from cool plasma at transition region temperatures \citep{Winebarger2013,Tiwari2019,Dolliou2023}.
Using the EM-loci technique, we also found a typical temperature of $\sim$10$^{5.5}$ K in the brightenings.
However, we cannot rule out the possibility that the brightenings are multi-thermal, containing million Kelvin plasma.
Because the cadence of the AIA 94, 131, and 335 passbands is 5 min for the observations taken on 2022 March 17, which is longer than the lifetime of the event, and the coronal loops overlap with the AFS in the AIA field of view,
we did not examine the intensity variation in AIA passbands of the event shown in \fig{fig:case3}.

Coordinated photospheric magnetograms taken by PHI are available for both datasets, and they have a similar spatial resolution to that of EUI. This characteristic helped us investigate the generation mechanisms of the bidirectional propagating brightenings.
All the examples are associated with flux emergence and/or cancellation, implying that they are triggered by magnetic reconnection.
\citet{Yadav2019} suggests the existence of small-scale loops beneath the large-scale loops of AFSs based on magnetic field extrapolation from magnetic field measurements in both the photosphere and chromosphere.
Magnetic reconnection may take place between the newly emerging small-scale loops and the preexisting large-scale loops in the AFSs and trigger the bidirectional propagating brightentings.
It is also possible that the event is triggered by small-scale reconnection between the neighboring flux tubes within the AFSs \citep[e.g.,][]{2020ApJ...899...19C,2021NatAs...5...54A,2021A&A...656L...7C}.
It is worth mentioning that the magnetic field is highly non-potential in filaments \citep[e.g.,][]{Gibson2018}, and scales of the corresponding magnetic elements approach the spatial resolution of PHI,
making it challenging to perform reasonable magnetic extrapolation for our study \citep[e.g.,][]{Wiegelmann2021}.

Magnetic reconnection can occur between the small-scale loops below AFSs near magnetic dips \citep{Mandrini2002}, so-called bald patches \citep{Titov1993}.
Ellerman bombs \citep{Ellerman1917,Georgoulis2002} or UV bursts \citep{Peter2014,Young2018} are often generated near bald patches, which often show clear compact emission enhancement in the 1600 {\AA} passbands \citep{Qiu2000,Vissers2019}.
In addition, magnetic reconnection sometimes occurs between the small-scale loops at higher altitude and produces loop-like brightenings in the vicinity of filaments \citep[e.g.,][]{Tiwari2014}.
Such processes are associated with flux cancellation in the photosphere, rearrange short loops into long loops, and generate sub-flares visible in the AIA 1600 {\AA} passband.
However, the events in our observations do not show significant emission enhancement in AIA 1600 {\AA} images, as shown in \fig{fig:aia}(i), and signatures of corresponding loop rearrangement.
This fact implies that the events are unlikely associated with magnetic reconnection among the small-scale loops.
Magnetic reconnection can also occur between AFSs and ambient coronal loops \citep{Huang2018},
and post-reconnection hot loops can remain bright in the AIA 94 {\AA} passband for a long time.
The long lifetime of the hot loops may be because of low densities and low radiative cooling rates in the loops or multiple heating episodes during the evolution of the loops \citep{Reale2019}.
However, the hot loops visible in the AIA 94 {\AA} channel in our observations are not connected to the loop-like brightenings within the AFS, as can be seen in \fig{fig:aia}.

\citet{Yan2018} studied the spectral observations of a loop within an AFS visible in AIA EUV and IRIS/SJI 1400 {\AA} images,
and they reported that the Si~{\sc{iv}} line profiles exhibit continuous blue-wing enhancements from $\sim$70 to $\sim$265 km~s$^{-1}$.
The loop is related to the flux cancellation in the photosphere underneath the AFS, and it is likely triggered by magnetic reconnection below or near the apex of the loops.
\citet{Yan2018} suggest that the emission enhancement of the Si~{\sc{iv}}, C~{\sc{ii}} lines is mainly due to density variation,
which may explain the faint SJI 1400 {\AA} emission enhancement of the event shown in \fig{fig:case3}.
Unfortunately, the IRIS slit did not cross any of the AFSs we investigated in our study.

Plasmoids, or blob-like structures, are often observed in various types of events triggered by magnetic reconnection on the Sun, including flares \citep[e.g.,][]{Takasao2012}, mini-filament eruptions \citep[e.g.,][]{Cheng2023,Li2023}, jets \citep[e.g.,][]{Kumar2018,Mandal2022}, and untangling of braided loops \citep{Chitta2022}.
In addition, numerical simulations have shown that magnetic reconnection between newly emerging magnetic flux and preexisting magnetic field structures can produce {localized brightenings and} plasmoids \citep[e.g.,][]{2017ApJ...839...22H,Ni2021}.
{In particular, \citet{2022ApJ...929..103T} reported bright dots within coronal bright points in EUI observations and reproduced them in MHD simulations. The bright dots therein were generated by magnetic reconnection between emerging flux and preexisting loops.}
The event observed on March 17 presents blob-like structures within the loop-like brightening, and the blobs propagate along the filament threads.
These blobs may be plasmoids generated during the magnetic reconnection processes through tearing instability.
Bidirectional propagating blob-like brightenings were also found in some filament channels.
\citet{Wei2023} reported bidirectional brightenings visible in different AIA EUV channels. These brightenings were triggered by internal reconnection and move along a small filament with speeds of 50--70 km~s$^{-1}$.
The brightenings are also visible in Solar Upper Transition Region Imager \citep[SUTRI;][]{SUTRI} Ne~{\sc{vii}} 465 {\AA} images, indicating the existence of plasma near 10$^{5.7}$ K. 
{It is worth mentioning that  \citet{2020ApJ...897L...2P} reported counter-streaming flows in a filament originating from jets triggered by flux cancellation at its end.}
%The localized magnetic reconnection is followed up by magnetic reconnection between the filament and ambient magnetic structures, which generates hot loops observed in AIA hot channels.
%
Such dynamic blobs have also been found in filament channels during their growing phase \citep{Li2022b},
and the small-scale reconnection is associated with flux cancellation in the photosphere, and it helped build up a twisted flux rope that later erupted.

%==============================================================================
\section{Conclusions\label{S:conclusions}}
%==============================================================================

In this study, we have investigated bidirectional propagating brightenings in two AFSs observed by EUI and PHI.
The propagating speeds of the brightenings reached 100--150 km~s$^{-1}$.
The brightenings may consist of many small-scale blobs, which can be plasmoids generated during magnetic reconnection.
The blobs are resolvable in EUI observations when Solar Orbiter is near perihelion.
The coordinated observations taken by SDO/AIA on 2022 March 4 show that the intensity variations of the brightenings in all EUV passbands match well, implying that the brightenings may be dominated by emissions at cool temperatures well below 1 MK.

Magnetograms taken by PHI have a similar spatial resolution to EUI, and the viewing angles of the two instruments are identical.
This makes it possible to study the magnetic field evolution in the photosphere beneath the brightenings.
We find that most events are associated with flux emergence in the photosphere.
Thus, we suggest that the newly emerging magnetic flux reconnects with preexisting magnetic field structures in the filament, and reconnection triggers the bidirectional flows propagating along the threads within the AFSs.
As bright lanes can be prevalent in AFSs, particularly in the one observed on March 4, such energy release driven by photospheric magnetic activities in the middle of the filaments may play an important role in mass and energy transfer within the AFSs.

%==============================================================================
%==============================================================================
%==============================================================================
%==============================================================================
\begin{acknowledgements}
{We thank the anonymous referee for constructive comments.}
The work of Y.C. and S.M. was supported by the Deutsches Zentrum f{\"u}r Luft und Raumfahrt (DLR; German Aerospace Center) by grant DLR-FKZ 50OU2201.
Y.C. acknowledges funding provided by the Alexander von Humboldt Foundation.
L.P.C. gratefully acknowledges funding by the European Union (ERC, ORIGIN, 101039844). Views and opinions expressed are however those of the author(s) only and do not necessarily reflect those of the European Union or the European Research Council. Neither the European Union nor the granting authority can be held responsible for them.
This research was supported by the International Space Science Institute (ISSI) in Bern, through ISSI International Team project {\#}23-586 (Novel Insights Into Bursts, Bombs, and Brightenings in the Solar Atmosphere from Solar Orbiter).
Solar Orbiter is a space mission of international collaboration between ESA and NASA, operated by ESA. The EUI instrument was built by CSL, IAS, MPS, MSSL/UCL, PMOD/WRC, ROB, LCF/IO with funding from the Belgian Federal Science Policy Office (BELSPO/PRODEX PEA 4000112292); the Centre National d’Etudes Spatiales (CNES); the UK Space Agency (UKSA); the Bundesministerium für Wirtschaft und Energie (BMWi) through the Deutsches Zentrum f\"ur Luft- und Raumfahrt (DLR); and the Swiss Space Office (SSO).
We are grateful to the ESA SOC and MOC teams for their support. The German contribution to SO/PHI is funded by the BMWi through DLR and by MPG central funds. The Spanish contribution is funded by AEI/MCIN/10.13039/501100011033/ and European Union “NextGenerationEU”/PRTR” (RTI2018-096886-C5, PID2021-125325OB-C5, PCI2022-135009-2, PCI2022-135029-2) and ERDF “A way of making Europe”; “Center of Excellence Severo Ochoa” awards to IAA-CSIC (SEV-2017-0709, CEX2021-001131-S); and a Ramón y Cajal fellowship awarded to DOS. The French contribution is funded by CNES.
IRIS is a NASA small explorer mission developed and operated by the Lockheed Martin Solar and Astrophysics Laboratory (LMSAL) with mission operations executed at NASA Ames Research center and major contributions to downlink communications funded by ESA and the Norwegian Space Centre. SDO is the first mission of NASA Living With a Star (LWS) Program.
\end{acknowledgements}

\bibliography{refs}{}
\bibliographystyle{aa}

\end{document}